# First-principles investigation of multifaceted properties; lattice dynamic, structural stability, mechanical, electronic, magnetic and thermodynamic response of Alkali metals-based semi Heusler alloys.

Diwaker[1], Shyam Lal Gupta[2], Anupam[3], Sumit Kumar[4], Aadil Fayaz[5] and Ashwani Kumar[6*]

[1]Department of Physics, SCVB Govt. Degree College Palampur-176061 (INDIA)
[2]Department of Physics, Harish Chandra Research Institute, Allahabad- 211019 (INDIA)
[3]Department of Physics, RGM Govt. Degree College Jogindernagar-175015 (INDIA)
[4]Department of Physics, Govt. Degree College Una-174303 (INDIA)
[5]Department of Physics, Panjab University Chandigarh-160014 (INDIA)
[*6]Department of Physics, Abhilashi University Mandi-175045 (INDIA

*E-mail-ashwani.bits05@gmail.com*

## Abstract

Taking into considerations the wide compositional stretch of Heusler alloys, the first-principles density functional theory-based calculations are excellently suitable for estimating the multifaceted properties of alkali metal based LiVSb and NaVSb Heusler alloys. We calculated ground state stability by optimizing the energy in α, β and γ phase configurations. The materials are dynamically stable in spin-polarised phase type-α. To explore the electronic structure, we successfully employed the generalised gradient approximation potential. The electronic band structures indicate a half-metallic nature featuring a wide indirect band gap of 1.40eV and 1.45eV. We computed the second-order elastic parameters at different pressure levels. The Pugh's ratio (< 0.25) assessed that both alloys are brittle in nature and mechanically stable. The obtained magnetic moment is consistent with the Slater-Pauling rule. By executing the Quasi-Harmonic Debye model and Boltzmann theory we assessed the various thermodynamic parameters and transport coefficients of both alloys at different temperatures and pressures. All positive frequencies in lattice dynamic study confirmed their stability. Our findings highlight the potential of these alloys in modern semiconductor technology, spintronics and thermoelectric applications.

## I. Introduction:

Half metals (HMs) are a novel class of materials which behave like a metal in one spin channel and for the other they act like a semiconductor [1, 2]. Consequently, HMs exhibit 100% spin polarised current [3]. The current due to charge carriers is being confined only in one spin channel which significantly enhances the competence of commercial spintronic devices like shape memory effect [4], magnetic tunnel junction (MTJ) [5], giant magneto-resistance (GMR) [6], opto-electronic, electro-optic devices [7] and high Curie temperature ($T_c$) [8]. Even though, the half- metals are abundantly proposed in many materials theoretically, among them only few are validated experimentally so far [9, 10]. Over the past decade Heusler alloys-a versatile class of HM have attracted a significant interest of researchers around the globe due to their adaptable chemical composition and robust magnetoresistive effect empowers them for various functionalities [11, 12]. The NiMnSb was the first half or semi Heusler which was predicted as HM [13] and later first full Heusler $Co_2MnZ$ (Z=Si, Ge) [14] was confirmed as half metal. Since then, a vast attention was given to explore them exhaustively. The literature reported other structures of the half-metallic magnets like Zn-blend type CrP and CrS [15], sp electrons ferromagnet with rock salt type crystal structures BaC and SrC [16], diluted magnetic semiconductors [17] and double perovskites such as $Sr_2FeReO_6$ and $MnFeReO_6$ [18]. Heusler alloys is found to be attractive class of ternary intermetallic alloys mainly categorized as half Heusler and full Heusler materials and denoted by stoichiometric formula XYZ and $X_2YZ$. Here X and Y are transition metal or alkali or alkaline Earth metal atoms whereas Z belongs to sp block element. In half Heusler alloys there occurs a possibility of leaving (face centred cubic) *fcc* sublattice vacant, whereas in full Heuslers there exist four interpenetrating *fcc* sublattices in full Heuslers. The half Heusler ternary alloys are new structural variant of their family. Alkali metals based half Heuslers are magnificent candidates for spintronics applicability due half-metallic characteristics, wide and stable band gap and large magnetisation [19-21]. Some cobalt based alloys are synthesis successfully but they are incompetent to attain high spin polarisation (SP) at normal temperature despites few exceptions and the estimated band gap for them is 0.2-0.5eV only [14]. They also exhibit low spin polarisation. The low spin polarisation at normal temperature is attributed generally to interface/surface and thermal fluctuations. However, materials with wider band gap openings are likely to keep half-metallicity even at elevated temperature and at interface or surface [22]. Meanwhile alkali metal (AM) based Heusler compounds are competent to fill this void. They exhibit much wider band than the transition metal (TM) based Heusler alloys [23, 20]. Comprehensive research is carried out to explore the various functionalities of lithium based Heusler alloys such as LiCrAs [24], LiMnZ (Z=N, P, Si) [25], LiBeAs, LiBeSb, LiBeBi, LiScGe, LiScSi [26] and LiCrAs [27] for the various useful fields of material science. Hopefully, this study extends a multifaceted understanding of Alkali metal based half Heusler alloys. Presently, the necessity of new and versatile materials is growing endlessly, while experimental breakthrough in new materials involve prolonged and expensive synthesis process meanwhile, computational methods emerged as most influential approaches to carve the problem. The main aim to pursue this work is to investigate structural phase stability, half-metallic, magnetic properties, lattice dynamic properties and thermodynamic response of Alkali metals based LiVSb and NaVSb semi Heusler alloys using the first-principle DFT calculations. We selected these alloys for their technological endurance and confirming that they are novel, cost effective and are not belongs to the category of rare elements. The proposed Alkali metals based LiVSb and

NaVSb semi Heusler compounds are not previously reported anywhere as per our knowledge. This paper is organized as follows: the section II concisely discusses the methodology adopted to accomplish the work. Section III comprehensibly presents and elaborates the results obtained from computational output pertaining to various properties, viz. structural, electronic, mechanical, lattice dynamics and thermodynamic performance. Finally, section IV reflects and summarise the work performed followed conclusions.

## II. Computational methodology

The present work is performed successfully and presented mainly in four key stages; first, to determine the most stable configuration by means of the volume optimisation and investigating the equivalent lattice parameters; second, computing the electronic band structures; third, to find the magnetic properties to check the magnetic moments value and fourth, to explore the lattice dynamic, mechanical and thermodynamic response of LiVSb and NaVSb alloys. To achieve first two, we have performed full potential linear augmented projector augmented plane wave (FP-LAPW) method [28] in density functional theory (DFT) executed in Wien2k simulation package [29]. The electron-electron interaction is effectively handled within the generalizes gradient approximation (GGA) and computed by Perdew-Burke-Ernzerhof (PBE) exchange-correlation potential [30]. In the LAPW approximation the Kohn-Sham equations [31] is simplified for a unit cell of two distinct basis sets for dissimilar regions, first region is non-overlapping atomic domain positioned at the atomic sites and the other is the interstitial region. To attains the last two steps as mentioned earlier, we utilised the semi classical Boltzmann transport equations using constant relaxation time approximation as executed in BoltzTrap code [32], IRelast [33] and Phonopy code [34]. To increase the spherical harmonics within the atomic sphere, $l_{max}$ is preferably set to 10. The value of $R_{Kmax}$ which is product of $RMT_{min}$ and $K_{max}$ is 8.0. Here, $RMT_{min}$ signifies the least muffin-tin radii of the atoms in the given unit cell and $K_{max}$ limits the interstitial plane wave of the basis sets. The structural dynamic stability and electronic band structure calculations were accomplished using a $10 \times 10 \times 10$ k-mesh in the 1$^{st}$ Brillouin zone. For the Brillouin zone sampling here, we utilized approximately 20,000 K-points. The convergence criteria for charge and energy were set to not as much smaller than 0.00001 Ry and 0.00001e/a.u.³, respectively. The convergence criteria are set at $10^{-4}$ Ry for energy and $10^{-4}$ eV for charge. The value of Seebeck coefficient (S) is computed without any adaptable constraints. However, the electronic part of electrical conductivity (σ) and the thermal conductivity (κ) needs to be calculated as functions of relaxation time (τ). Hence, instead of computing the absolute power factor S²σ, we have preferably calculated the figure of merit $T = \frac{S^2 \sigma T}{\kappa}$ without any adjustable constraints. All the thermoelectric parameters were computed as a function of chemical potential $(\mu - \varepsilon_F)$ at five different temperatures; 250 K, 500 K, 750 K, 1000K and 1250K respectively.

## III. Results and discussions

### A. Structural model

The half/semi and full/regular Heusler alloys are mainly crystallize in non-centrosymmetric cubic *fcc* in structure type C1$_b$ and *f-43m* with space group no. #216, whereas regular Heusler relaxed in centrosymmetric cubic *fcc* with L2$_1$ structure and *fm-3m* space group no. #225 respectively. The crystal structure of α-phase half Heusler AYZ (A=Li, Na; Y=V; Z=Sb) system is based on the rock salt (RS) sublattice of YZ. In AYZ type configuration AY or YZ forms the sublattices. The three possible phase types α, β and γ-phases and their corresponding Wyckoff's positions are shown in the **Table 1**.

**Table 1:** Wyckoff's positions for AVSb (A=Li, Na) half Heusler alloys in α, β and γ-phases

| Structure | Li/Na | V | Sb |
|---|---|---|---|
| α | (0.25, 0.25, 0.25) | (0.5, 0.5, 0.5) | (0, 0, 0) |
| β | (0.5, 0.5, 0.5) | (0, 0, 0) | (0.25, 0.25, 0.25) |
| γ | (0, 0, 0) | (0.25, 0.25, 0.25) | (0.5, 0.5, 0.5) |

The computed detail validates that among all β-phase is likely to be the most stable one. The magnetic ordered crystal structure for AVSb (A=Li, Na) is given in schematic **Fig. 1 (a) & (b).**

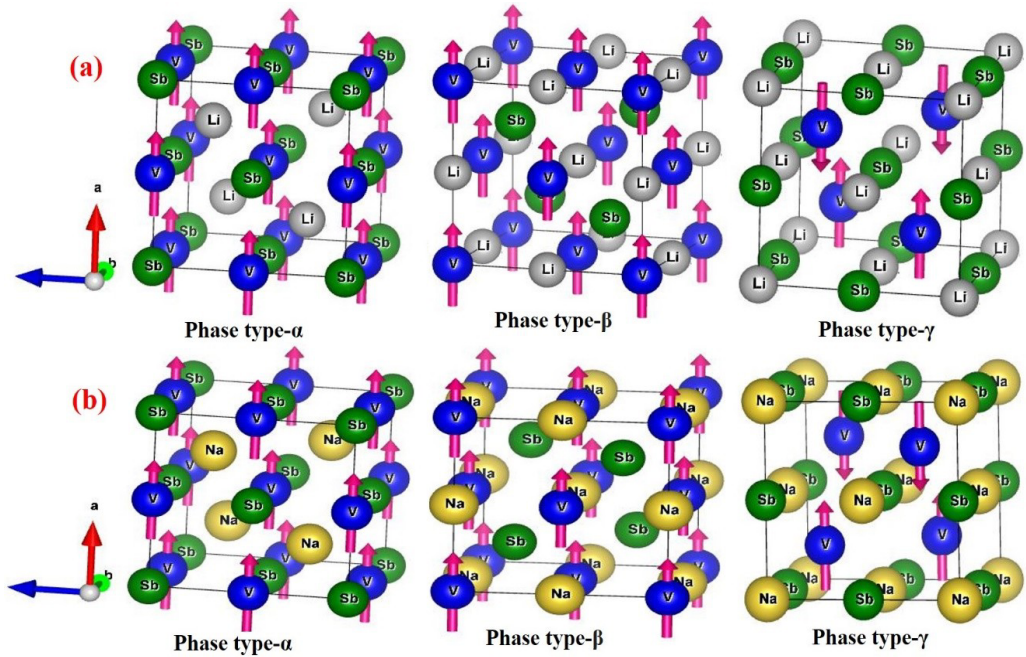

**Fig. 1:** Crystal structures in C1$_b$ symmetry for (a) LiVSb and (b) NaVSb alkali metal-based semi Heusler alloys in phase type-α, β and γ

The semi-Heusler AVSb (A=Li, Na) alloys display a structure C1$_b$ with three different atomic arrangements known as to α, β and γ phases. The α phase of both alloys are dynamically more stable ones. From the volume energy diagram as shown in **Fig. 2 (a) & (b)** we may predict that both AVSb (A=Li, Na) alloys with space group *F-43m* (#216) are dynamically more stable in the spin polarised (SP) phase than the non-magnetic (NM).

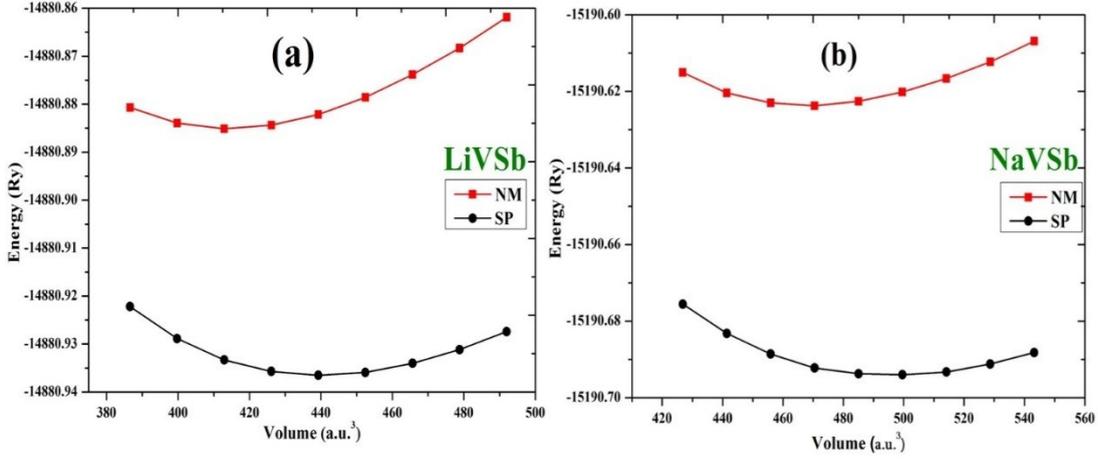

**Fig. 2:** Volume optimization curves in (a) spin polarized (SP) and (b) non-magnetic (NM) with stable phase type-α for AVSb (A=Li, Na) semi Heusler alloys

The distinguished lattice parameters are given in **Table 2**, here we observed that the mean lattice parameters and least energy for the cubic AVSb (A=Li, Na) system is inherited by stable type-I crystal structure. For the volume optimisation in spin polarized (SP) state Murnaghan eqn. of states [35] is implemented effectively and given in equation (1).

$$E_T(V) = E_0 + \frac{B_0 V}{B_0'(B_0'-1)}\left\{\left(\frac{V_0}{V}\right)^{B_0'} - 1\right| + B_0'\left(1 - \frac{V_0}{V}\right)\right\} \quad (1)$$

Here, $E_\theta$ is the total energy, $V_0$ is the volume of the unit cell at zero pressure and $B_0$ signifies the bulk modulus respectively. The computed lattice constant ($a_0$) at symmetry with GGA for both LiVSb and NaVSb semi Heusler alloys are 6.3874Å and 6.6475Å respectively. The other parameters like $V_0$, $B_0$, $V_0$ and $E_0$ reported first ever for AVSb (A=Li, Na) system are described in **Table 2**.

**Table 2:** Computed structural dynamic parameters for AVSb (Li, Na) semi Heusler alloys

| Alloy | Phase Type | $a_0$ (Å) SP | $a_0$ (Å) NM | B (GPa) SP | B (GPa) NM | Volume (Å) SP | Volume (Å) NM | Energy (Ry) SP | Energy (Ry) NM |
|---|---|---|---|---|---|---|---|---|---|
| LiVSb | α | 6.3874 | 6.2608 | 52.3714 | 62.9883 | 414.0266 | 439.6595 | -14880.9365 | -14880.8850 |
|  | β | 6.3165 | 6.3234 | 52.4356 | 63.7432 | 415.1601 | 438.6754 | -14880.6654 | -14880.8005 |
|  | γ | 6.3671 | 6.5634 | 51.9803 | 62.5439 | 416.2134 | 436.7653 | -14880.5986 | -14880.7901 |
| NaVSb | α | 6.6475 | 6.5230 | 44.4850 | 56.9710 | 495.5780 | 468.2505 | -15190.6942 | -15190.6223 |
|  | β | 6.5647 | 6.5002 | 43.7681 | 55.6543 | 496.6579 | 467.5431 | -15190.6212 | -15190.5687 |
|  | γ | 6.5693 | 6.4809 | 44.5645 | 57.9876 | 494.6543 | 469.5365 | -15190.6008 | -15190.5864 |

**B. Mechanical properties**

The elastic constants are generally derived from the second derivatives of the total energy, are vital for studying the stability of crystal structures, nature of chemical bonding and their mechanical response under the external forces. It offers insights into deformation attains by materials under certain exterior force and then to gain the original configurations. In the present investigations we have determined the various elastic constants at pressure from 5GPa to 10GPa. These calculations involved assessing the total elastic tolerance as a function of external strains. The AVSb (A=Li, Na) semi Heusler compounds in phase type-α shows stability in C1$_b$ cubic symmetry. Consequently, we computed the values of three independent elastic parameters $C_{11}$, $C_{12}$, and $C_{44}$ which are specific to the cubic system only. The parameter $C_{11}$ is attributed to longitudinal expansion and $C_{12}$ quantifies the transverse deformation. To discover the mechanical stability, equilibrium elastic coefficients must satisfy the. On the basis of computed results as given in **Table 3**, both LiVSb and NaVSb compounds confirmed the mechanical stability in accordance with the Born's stability criterion for cubic system, ($C_{11}$- $C_{12}$) > 0, $C_{11}$, $C_{12}$ and $C_{44}$ > 0, and ($C_{11}$+2$C_{12}$) > 0 and $C_{12}$< $B_s$ < $C_{11}$. Here we see that $C_{12}$ is less than $C_{44}$, suggesting the presence of covalent bonds and the angular dependency of interatomic forces is turned essential. The elastic moduli in both compounds are somewhat similar. Furthermore, a thorough analysis of elastic parameters provides insights into stability and phase transitions in the given system. The detail given in **Table 3** illustrated that AVSb (A=Li, Na) semi Heusler compounds fulfil the stability conditions and signifying its robust mechanical stability. Furthermore, the other parameters like and bulk modulus (B) and shear modulus (G) regulates the stiffness and compressibility respectively. The resistance to the plastic deformation, a force against the fracture, can be known by calculating B and G as follows-

$$B = \frac{C_{11} + 2C_{12}}{3} \tag{1}$$

and
$$G = \frac{G_R + G_v}{2} \tag{2}$$

The subscripts in Eqn. 2 i.e. R and V denotes the Reuss and Voigt bounds respectively. The values of B+G$_v$ and G$_R$ is resolute by executing the following relations-

$$G_R = \frac{5C_{11}(C_{11} - C_{12})}{5C_{11} + 3(C_{11} - C_{12})} \tag{3}$$

Here
$$G_v = \frac{C_{11} - C_{12} + 3C_{44}}{5} \tag{4}$$

The mathematical correlation between the various elastic constants is in accordance with the cubic crystal symmetry [36]. The internal strains pertaining to bond elongation and bond twisting are calculated by using Kleinmann parameter ($\xi$. It assesses the simple bond bending under external forces. The computed value of $\xi$ show that LiVSb and NaVSb alloys are more resistant to wider range of forces and hence validate them for industrial applications. To check the thermodynamic stability, specific heat at low temperature, phonon stability, we calculated the Debye temperature $(\theta_D)$ by means of the average sound velocity $(v_m)$-

$$\theta_D = \frac{h}{k} \left[ \frac{3n}{4\pi} \left( \frac{\rho NA}{M} \right) \right]^{1/3} V_m \tag{5}$$

Here, average sound velocity $(V_m)$ is find out by using the longitudinal $(V_l)$ and transverse $(V_T)$ velocity in the form shown as-

$$V_m = \frac{1}{3} \left( \frac{2}{V_s^3} + \frac{1}{V_l^3} \right)^{-1/3} \tag{6}$$

The calculated value of Debye temperature $(\theta_D)$ at 5GPa pressure is 1401.94K and 1431.338K for LiVSb and NaVSb alloys. To determine the melting temperature, we used the Fine's relation [37] and elastic constant-

$$T_m(K) = [553(K) + (5.911)C_{11}] GPa \pm 300 K \tag{7}$$

For LiVSb and NaVSb the calculated values of melting temperature at 5GPa are 6187K and 7095K respectively. High values of melting temperature confirms that both materials are able to maintain their ground state structure over a wide range of temperature and hence leads towards the stability.

The optimised values of the other essential elastic parameters are given in **Table 3-**

**Table 3:** Computed values of elastic parameters ($C_{11}$, $C_{12}$, $C_{44}$ in GPa), Voigt, Reuss & Hill's bulk, shear and Young's modulus ($B_v$, $B_R$, $B_H$, $G_v$, $G_R$, $G_H$ & $E_v$, $E_R$, $E_H$ in GPa),Voigt, Reuss & Hill Poisson coefficients ($V_v$, $V_R$, $V_H$ in GPa), Pugh's ratio (k), Kleinman's parameter ($\xi$), Ranganathan and Kube Anisotropy Index ($A_R$ & $A_K$), transverse, longitudinal and average wave velocity $(V_t, V_l \wedge V_a \in m/s)$, Debye temperature $(\theta_D \in K)$,Chen-Vickers & Tian-Vickers hardness $(H^{CV}, H^{TV} \in GPa)$, Lame's 1$^{st}$ & 2$^{nd}$ order parameters ($\lambda$, $\mu$ in GPa) under external pressure of 5GPa and 10GPa for cubic LiVSb and NaVSb alloys

| Stress Parameters | LiVSb | | NaVSb | |
|---|---|---|---|---|
| | 5GPa | 10GPa | 5GPa | 10GPa |
| $C_{11}$ | 953.232 | 1077.862 | 1106.816 | 1471.827 |
| $C_{12}$ | -448.034 | -414.328 | -501.922 | -635.780 |
| $C_{11}$-$C_{12}$ | 1401.266 | 1492.190 | 1608.738 | 2107.608 |
| $C_{11}+2C_{12}$ | 57.164 | 249.206 | 102.972 | 200.267 |
| $C_{44}$ | 774.060 | 932.721 | 818.005 | 1057.690 |
| $B_v$ | 19.055 | 83.069 | 34.324 | 66.756 |
| $B_R$ | 19.055 | 83.069 | 34.324 | 66.756 |
| $B_H$ | 19.055 | 83.069 | 34.324 | 66.756 |
| $G_v = (C_{11} - C_{12} + 3C_{44})/5$ | 744.689 | 858.071 | 812.551 | 1056.135 |
| $G_R = 5C_{44}(C_{11} - C_{12})/[4C_{44} + 3(C_{11}-C_{12})]$ | 742.917 | 847.886 | 812.495 | 1056.132 |
| $G_H$ | 743.803 | 852.978 | 812.523 | 1056.134 |
| $E_v$ | 159.266 | 579.358 | 274.171 | 505.035 |
| $E_R$ | 159.239 | 577.796 | 274.169 | 505.035 |
| $E_H$ | 159.253 | 578.580 | 274.170 | 505.035 |
| $v_v$ | -0.893 | -0.662 | -0.831 | -0.761 |
| $v_R$ | -0.893 | -0.659 | -0.831 | -0.761 |
| $v_H$ | -0.893 | -0.661 | -0.831 | -0.761 |
| $\xi = C_{11} + 8C_{12}/7C_{11} + 2C_{12}$ | -0.348 | -0.267 | -0.332 | -0.312 |
| $A_R$ | 0.012 | 0.060 | 0.000 | 0.000 |
| $A_K$ | 0.005 | 0.027 | 0.000 | 0.000 |
| $V_t$ | 12242.652 | 12722.871 | 12936.318 | 14259.757 |
| $V_l$ | 14271.757 | 15218.171 | 15172.360 | 16851.515 |
| $V_a$ | 12789.765 | 13371.361 | 13534.021 | 14945.093 |
| $\theta_D$ | 1401.941 | 1495.304 | 1431.338 | 1616.494 |
| k | 0.026 | 0.097 | 0.042 | 0.063 |
| $H^{CV}$ | 6960.649 | 1578.556 | 4081.529 | 2968.703 |
| $H^{TV}$ | 6401.268 | 1545.141 | 3858.867 | 2938.368 |
| $\lambda$ | -476.814 | -485.584 | -507.358 | -637.333 |

| | μ | 743.803 | 852.978 | 812.523 | 1056.134 |

The Pugh ratio (B/G) is a prominent parameter that describes the brittle or ductile nature of materials. Materials with a B/G ratio less than 1.75 are generally brittle. The calculated value of B/G ratio for both alloys LiVSb and NaVSb prove them as brittle material.

## C. Electronic properties

The electronic band structures of LiVSb and NaVSb exhibiting the half-metallic nature is shown in **Fig. 3**. We observed a relatively wide band gap (BG) unlike the previously reported conventional transition metal half-Heusler compounds. Both LiVSb and NaVSb are stabilise in type-α phase. In **Fig. 3**, the horizontal red lines represent the valence band maxima (VBM) and conduction band minima (CBM) and width between them is shown by an arrow suggesting the indirect band gap in spin down configuration. The band gap is shown by minority spin bands however, majority spin bands are metallic consequently, exhibit a half-metallic property. The calculated band gap is 1.405eV and 1.473eV for LiVSb and NaVSb respectively. The origin of half-metallic ferromagnetism (HMF) in these materials can be understood by investigating the projected density of states (pDOS) of all the metal atoms, as shown in **Fig. 4**. The pDOS for LiVSb and NaVSb is illustrated as the orbital fillings in the V and Sb elements play the prominent role.

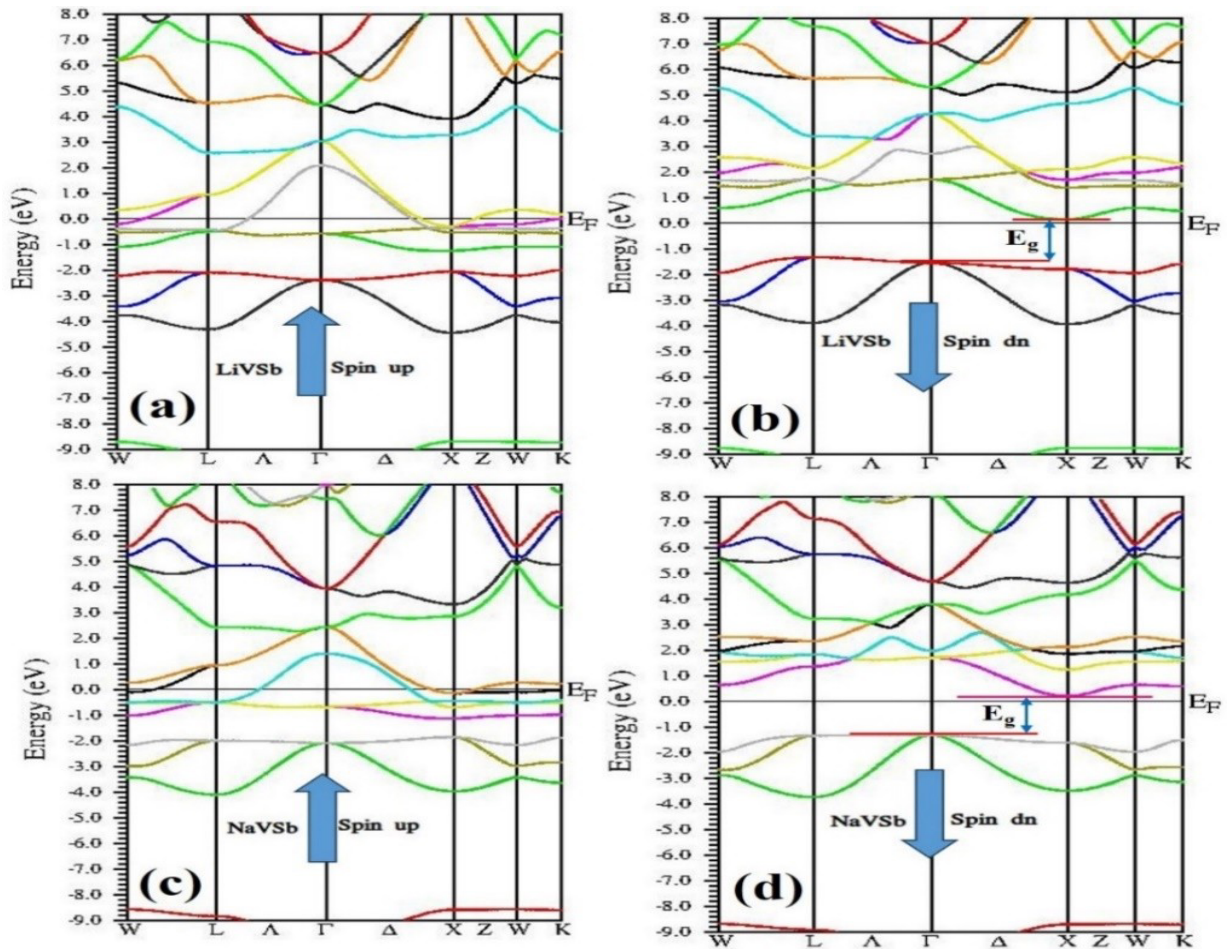

**Fig. 3:** Electronic band structure of LiVSb and NaVSb Heusler alloys for spin up and spin down configuration.

According to **Fig. 1(a)** and **(b)** and the discussion in reference [38], Sb atoms play as ligand anions. Each V atom is bounded by four Sb atoms in tetrahedral crystal structure. In this structure, the Sb atoms proceed towards the V atom and apply a repulsive force on the inter-axial d-orbitals of V. This force more or less destabilizes the inter-axial orbitals and results rise in their energy however, the axial orbitals keep on more stable as they are far from the ligands. This results in a significant splitting in energy level. Due to the ligand-field effect the d-orbitals of V to split into two subgroups one is $E_g$ and other is $T_{2g}$ state, here $E_g$ orbitals state possess lower energy than the $T_{2g}$ state. The s orbital of Li and Na atom are separated from other one by large energy gap and hence ensure negligible contribution around the Fermi energy domain. Contrary to this the Sb–p and V–d orbitals inhabit around the $E_F$ (check **Fig. 4**). The robust exchange splitting between the V–d orbitals contribute to the HM gap and a large localized magnetic moment at their site. This band gap arises from the electronegativity imbalance between the highly electropositive Na/Li atoms and highly electronegative Sb atom. Explicitly, the Li/Na atoms in LiVSb and NaVSb system transfers its s-state electron to its p-states, which subsequently contribute to the [VSb]⁻ states. Understanding the electronic band profile is crucial for determining the suitability of materials in the fabrication of various solid-state devices. The band profile of LiVSb and NaVSb Heusler alloys is explored by calculating the projected (pDOS) and total density of state (TDOS). The TDOS and pDOS of LiVSb and NaVSb at normal temperature and pressure are shown in **Fig. 4 (a to f)**. It is observed that in spin down configuration there is a wide band gap between conduction and valence band for both the materials. The peak occurs near the Fermi level is due to the interaction between d state electrons of V atoms and p electrons of Sb atoms whereas, the contribution from Li and Na atoms is almost negligible. The detailed information about contribution from individual atoms is given in pDOS profile of both alloys.

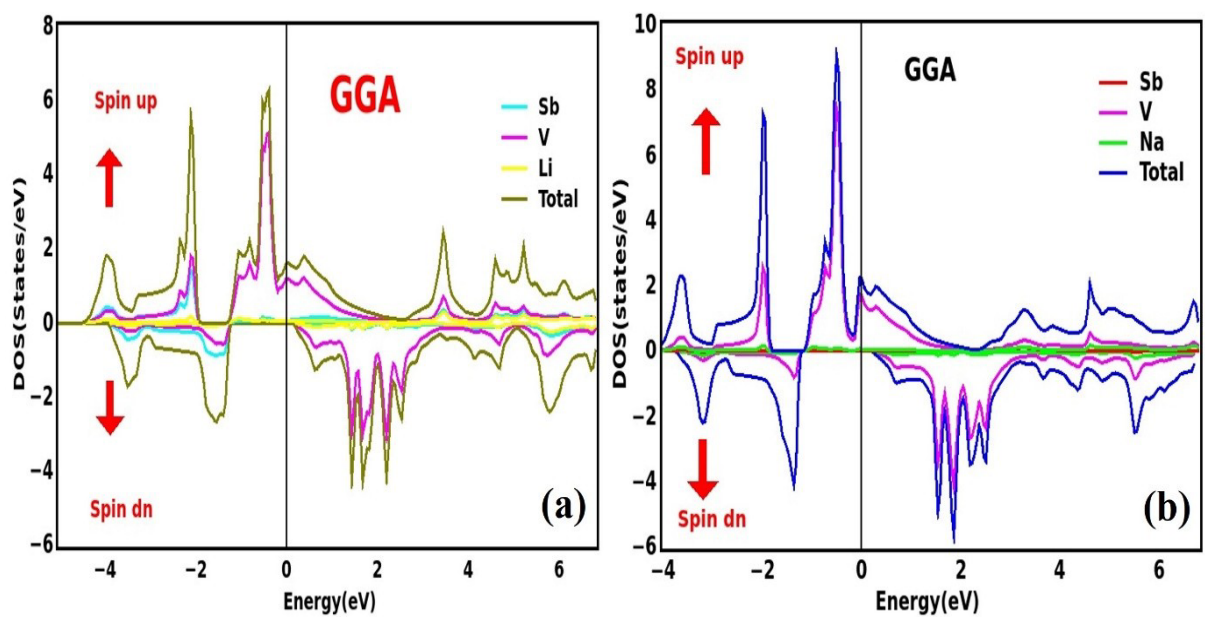

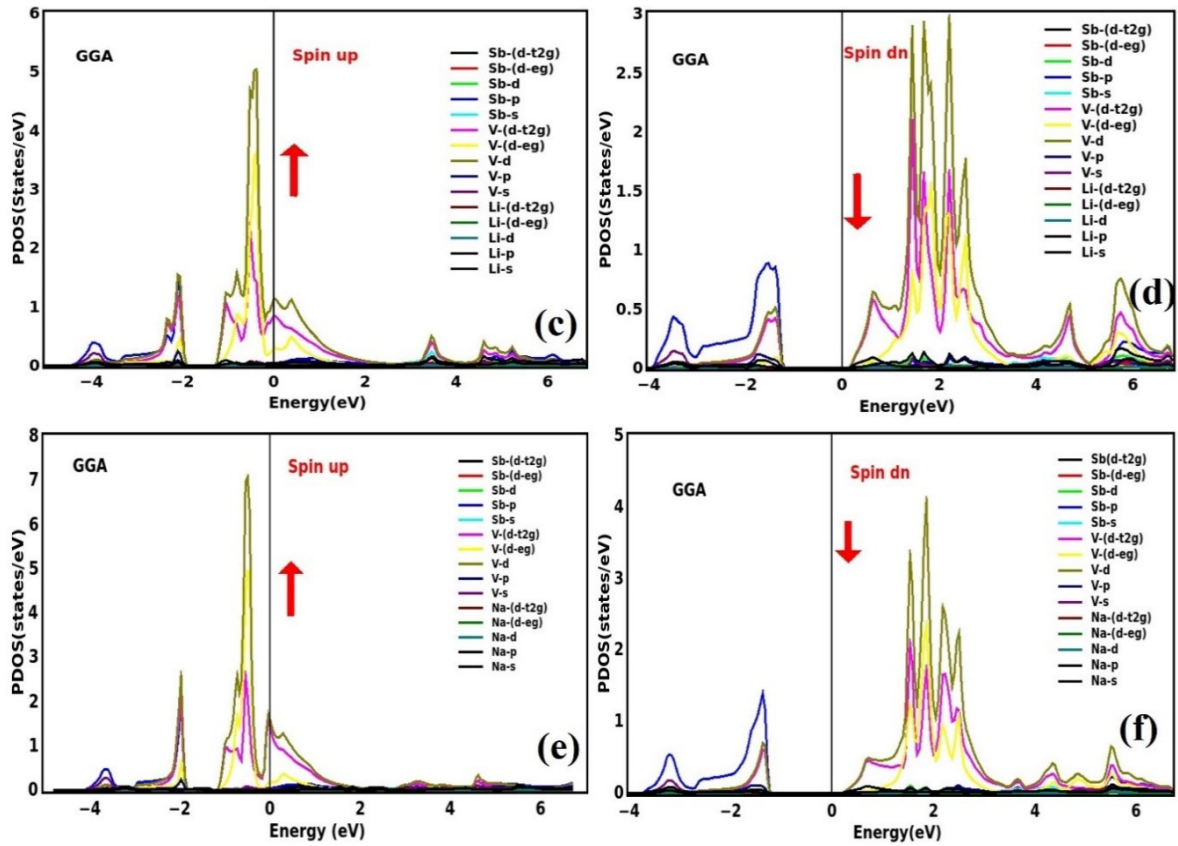

**Fig. 4:** TDOS and pDOS plots from (a) to (f) for AVSb (A=Li, Sb) semi Heusler system

We see that the covalent character of the tetrahedral sublattice facilitate to define the chemical bonding between neighbouring atomic orbitals using the molecular orbital theory. In case of Vanadium (V), there is a ligand field effect which causes the $d$-orbitals to split into two corresponding subgroups known as $e_g$ and $t_{2g}$ states. Here the $e_g$ orbitals have lower energy as compared to the $t_{2g}$ orbitals states. Furthermore, the s state of sodium (Na) atom is adequately separated from other states which renders them to play insignificant role near the Fermi region ($E_F$). However, the Sb–p and V–d orbitals be located around the Fermi level as seen in **Fig. 5**. The important exchange splitting of vanadium (V) d-orbitals results in a band gap which is indicative of substantial localized magnetic moment (MM) and half-metallicity at the V site.

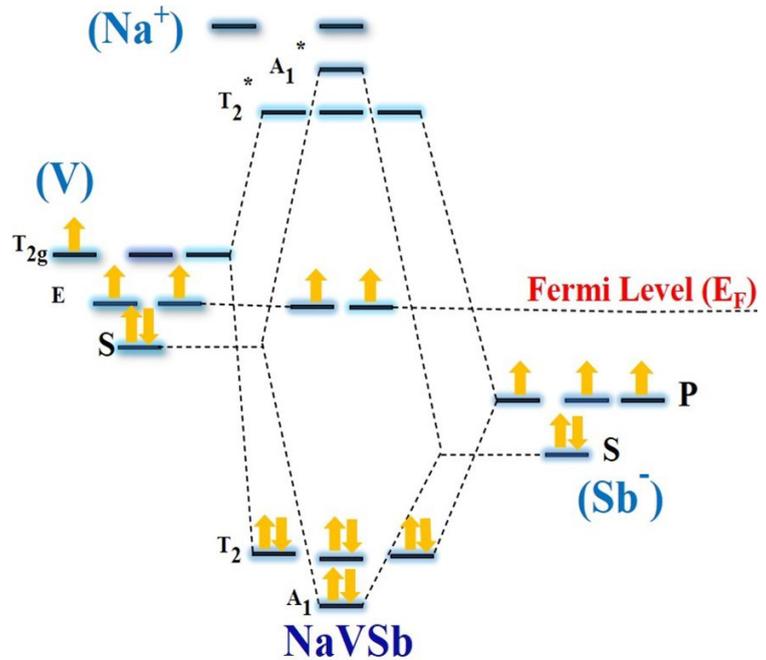

**Fig. 5:** Ligand field splitting in NaVSb system

The band gap (BG) in both LiVSb and NaVSb alloys arises from the electronegativity differences between the atoms. The lithium (Li) and sodium (Na) metal atoms being the most electropositive, and antimony (Sb) being the most electronegative contribute to this effect. The Na atom encourages its *s*-state electron to its *p*-states, which contributes and form the [VSb]$^-$ states, indicating that Na effectively donates its electron.

The 4p orbitals of vanadium (V) are empty, and its 3d orbitals are partially filled. These 3d orbitals splits up into triply degenerate $T_{2g}$ and doubly degenerate $E_g$ orbitals. The 5s and 5p orbitals of antimony (Sb) hybridize with the s and $T_{2g}$ orbital states of V, making a set of bonding $A_1$, triply degenerate $T_2$ orbitals and a set of anti-bonding $A_1^*$ as well as triply degenerate $T_2$ orbitals. The $E_g$ orbitals remains non-bonding. The vacant 2s orbital of Na metal atom is resides above the anti-bonding orbitals of [VSb]$^-$. Based upon above explanations we draw the crystal field diagram of the predicted alloys in **Fig. 5**. We observed that eight electrons hold the bonding states whereas only two valence electrons will reside in non-bonding hybrid orbitals. These electrons aligned their spins in accordance with Hund's rule, resulting in a magnetic moment of 3μ$_B$. The Fermi level (E$_F$) is lying in the gap among the bonding and anti-bonding orbitals in addition to spin-up non-bonding states and leads genesis of half-metallicity in alloys.

The lowest bands arise from the bonding $A_1$ states and the next bands originates from the triply bonding $T_2$ states and non-bonding $E_g$ as well as anti-bonding $T_2^*$ orbitals states. In the majority spin states the non-bonding $E_g$ states lies on the Fermi level (E$_F$). Hence there exist an indirect gap of 1.405 eV and 1.473eV system between the Γ and X point for the minority spin channel. for LiVSb and NaVSb alloys respectively.

**D. Magnetic Properties**

Given **Fig. 4,** illustrates that the total density of states (TDOS) for the minority and majority spin states are counterbalanced and this leads to arise of significant magnetic moment (MM) in the given alloys. The computed magnitude of MM for a half-metal is an integer value and also it is in accordance with the Slater-Pauling (S-P) rule expressed as $M_{total} = Z_{total} - 8$ [39]. Here $M_{total}$ is the total MM and $Z_{total}$ signifies the total number of electrons and can be calculated by adding the number of valence electrons in the constituent atoms. For the present system the calculated values of magnetic moment for LiVSb and NaVSb alloys are 3.00μ$_B$ and 2.99μ$_B$ respectively and found consistent with S-P behaviour of half-metallic ferromagnets. The **Fig. 6** explains the intrinsic and total magnetic moment as observed in LiVSb and NaVSb materials. There is negligible contribution from Sb metal atom to define the MM. The d-block metal atom, vanadium (V), is the primary contributor, whereas the participation of lithium (Li) and sodium (Na) atoms in total magnetic moment is minimal.

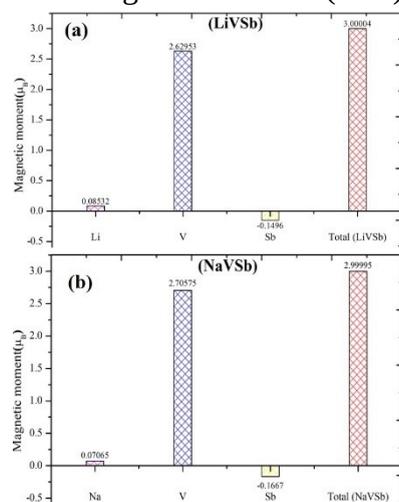

**Fig. 6:** Intrinsic and total magnetic moments for (a) LiVSb and (b) NaVSb alloys

The electronic band profile, TDOS, pDOS and obtained magnetization value all show that the alloys exhibit the half-metallic (HM) behavior. Another important property for spin injectors, especially from an application perspective is the Curie temperature (TC). Curie temperature (Tc) is an essential property of magnetic solids. It is the temperature at which the solid experiences a phase transition from magnetic to non-magnetic. At this temperature, the intrinsic MM values significantly change their orientations. The value of Tc is obtained by executing MM values in a linear relationship [40]-

$$T_c = 23 + 181\, M_{total} \tag{8}$$

$T_c$ is the Curie temperature here and $M_{total}$ shows the total MM. The computed value of Tc for LiVSb and NaVSb alloys is 566K and 565K respectively. The results validating these materials as potential candidate for high temperature applications.

**E. Thermodynamic and thermoelectric response**

In this section thermodynamic properties are thoroughly investigated to know the role of LiVSb and NaVSb alloys for thermoelectric technology. The thermoelectric materials transform waste heat into usable energy and present a way to create efficient and environmentally friendly energy sources. This requires the material to be optimally and efficiently transmit both heat as well as charge within it. Semi or Half Heusler alloys are the potential candidates for their transport behaviour. To demonstrate the thermoelectric competency of these material's, the fluctuations of significant transport parameters is drawn at a temperature scale from 50 to 600K. The Quasi-Harmonic Debye model [41] is executed to know the thermodynamic stability of both alloys. The Gibbs function is taken as –

$$G(V:P,T) = E(V) + A_{Vib}[\theta(V):T] \tag{9}$$

Here E(V) signifies the energy, ѳ(V) is the Debye temperature and $A_{Vib}$ denotes the vibrational energy term. The volume expansion behaviour, thermal expansion coefficient (α), specific heat at constant temperature ($C_P$) and pressure ($C_V$) and entropy (S) are involved in the computed thermodynamic potentials in the temperature and pressure range of 0–600K and 0–10GPa respectively. The specific heat at constant volume and ($C_v$) and constant pressure ($C_p$) are the crucial parameters to predict the intensity of phase transitions in lattice vibrations which occur in the materials under different temperature and pressure. The higher atomic or molecular mobility suggests the high-temperature instability. The variation of temperature with volume (V), thermal expansion (γ) and entropy is presented in **Fig. 7 (a) to (f)** for LiVSb and NaVSb. The volume decreases considerably with pressure and increase with temperature. The thermodynamic eqn. of state is known by computing the variation of thermal expansion coefficient (α) with temperature (T). **Fig. 7 (c)** and **(d)** shows the variation of α with T, there is a rapid increase in α at lower temperatures and ultimately becomes constant at higher range due to the suppression of anharmonic effects. The magnitude of α gradually decreases with increase in pressure from 0 to 10 GPa.

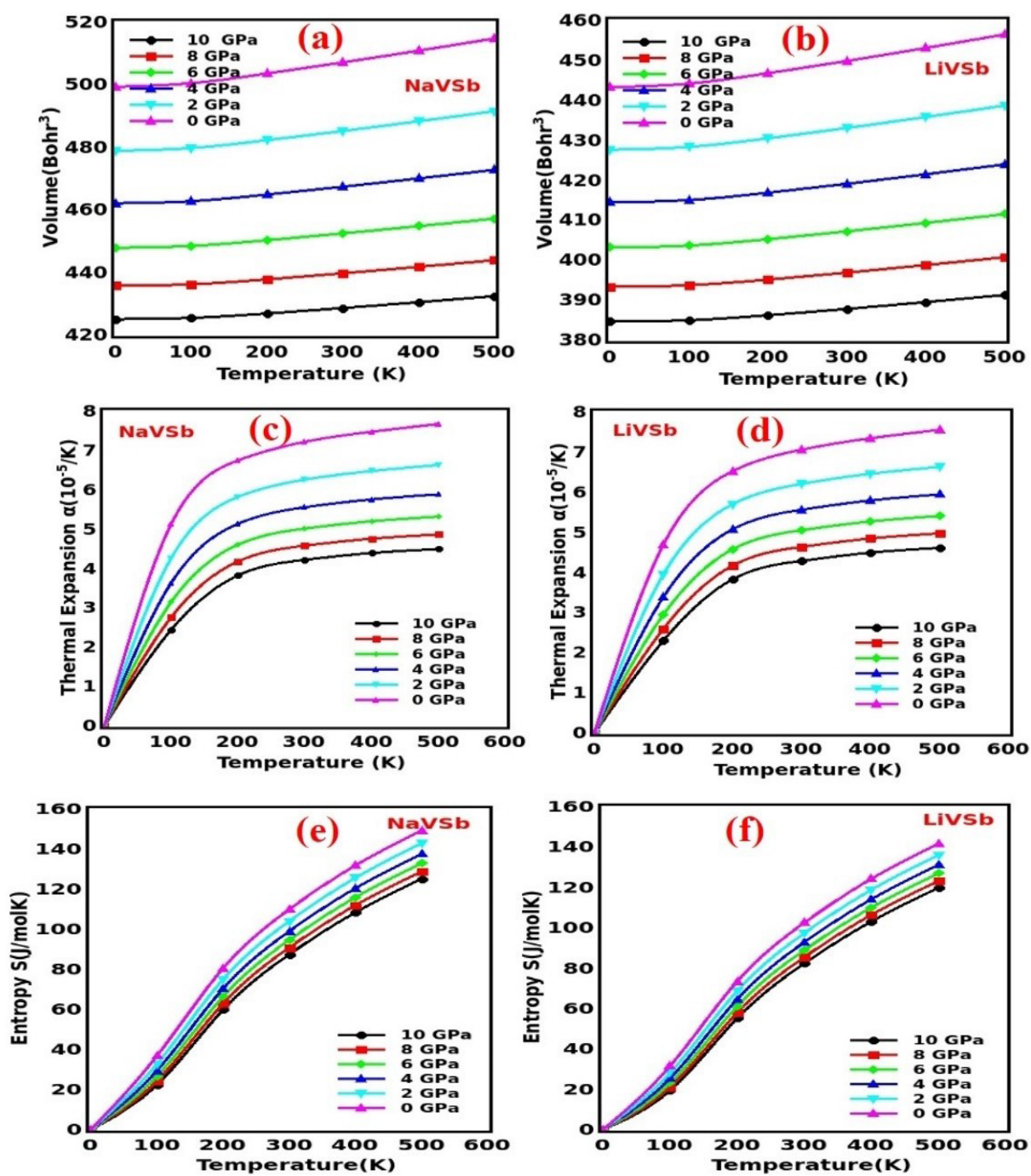

**Fig. 7:** Variation of V, α and S with temperature from **(a)** to **(f)** for LiVSb and NaVSb alloys

The values of various thermodynamic parameters at 0GPa and 300 K are highlighted in the **Table 4**.

**Table 4:** The estimated value of thermodynamic parameters: thermal expansion (α) in ($10^{-5}$ $K^{-1}$), specific heat at constant volume ($C_V$) and constant pressure ($C_p$) in (J $mol^{-1}$ $K^{-1}$) and entropy (S) in ($J^{-1}$molK) at 0 GPa pressure and room temperature

| Alloys | α | $C_v$ | $C_p$ | S |
|--------|-----|------|------|-----|
| LiVSb | 6.8 | 65.7 | 69.5 | 100 |
| NaVSb | 7.0 | 70.5 | 70.4 | 110 |

The computed values of $C_V$ and $C_P$ for LiVSb and NaVSb are given in **Fig. 8 (a)** to **(d)** which helps us to know the effect of the increase in atomic vibrations due to the heat absorption. The plots illustrated that for different pressure values, $C_V$ and $C_P$ increases exponentially at low temperatures, although at higher temperatures it achieves a constant value and follows the Dulong-petit law.

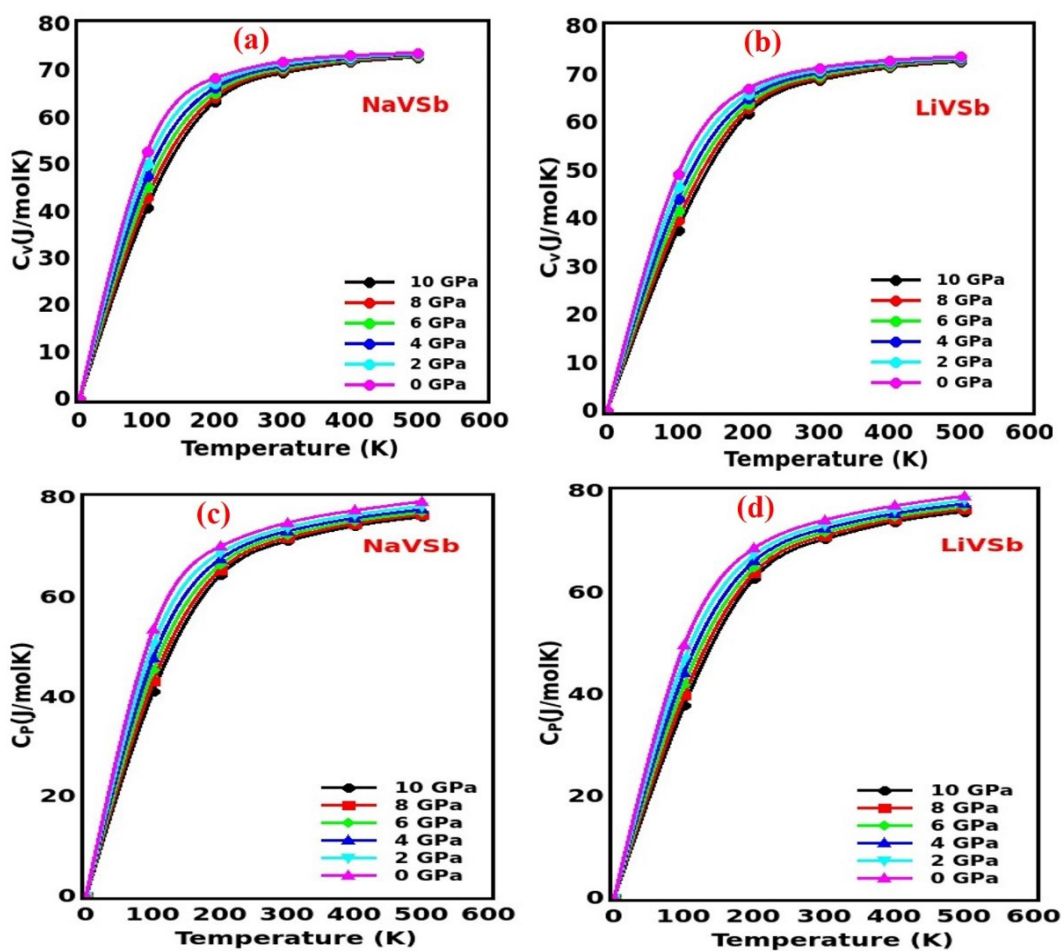

**Fig. 8:** Variation of specific heat at constant volume ($C_V$) and constant pressure (CP) from **(a)** to **(d)** for LiVSb and NaVSb alloys

The fluctuations of other outstanding transport parameters are plotted from 0 to 600K to determine the thermoelectric performance of materials. The various distinguished thermoelectric coefficients are shown as follows-

$$\sigma = e^2 \int x(\varepsilon) \left( \frac{-\partial f_0}{\partial \varepsilon} \right) \partial \varepsilon \qquad (10)$$

$$S = \frac{e}{T\sigma} \int x(\varepsilon) \left( \frac{-\partial f_0}{\partial \varepsilon} \right) (\varepsilon - \mu) \partial \varepsilon \qquad (11)$$

$$\kappa = \frac{1}{T} \int x(\varepsilon) \left( \frac{-\partial f_0}{\partial \varepsilon} \right) (\varepsilon - \mu)^2 \partial \varepsilon \qquad (12)$$

The Seebeck coefficient (S) and figure of merit (zT) imparts a clear information about the efficacy of material's performance. In real sense the electronic structure near the Fermi level ($E_F$) provides the foundation for the efficiency of thermoelectric materials. **Fig. 9 (a)** signifies how the Seebeck coefficient (S) changes with temperature for LiVSb and NaVSb. The negative values of S determines that the majority carriers are electrons in both materials. **Fig. 9 (b)** represent the schematic diagram for figure of merit (zT) as a function of temperature (T). At temperature of 1200 K, the obtained zT value is 0.0060 and 0.0065 for LiVSb and NaVSb alloys respectively. Another important thermoelectric (TE) parameter is the power factor (PF), which estimate the quantity of electrical energy created. A solid with high PF values is considered as ideal candidate for TE applications. **Fig. 9 (c)** depicts the PF for these alloys is in the temperature range from 200K-1200K. The PF is showing the elevated trends from minimum to the optimum values. The rise in electrical conductivity leads to increase in the PF of TE materials. This is due to reason that PF and electrical conductivity are linked with each other via a linear relationship *i.e.* $PF = S^2 \sigma$.

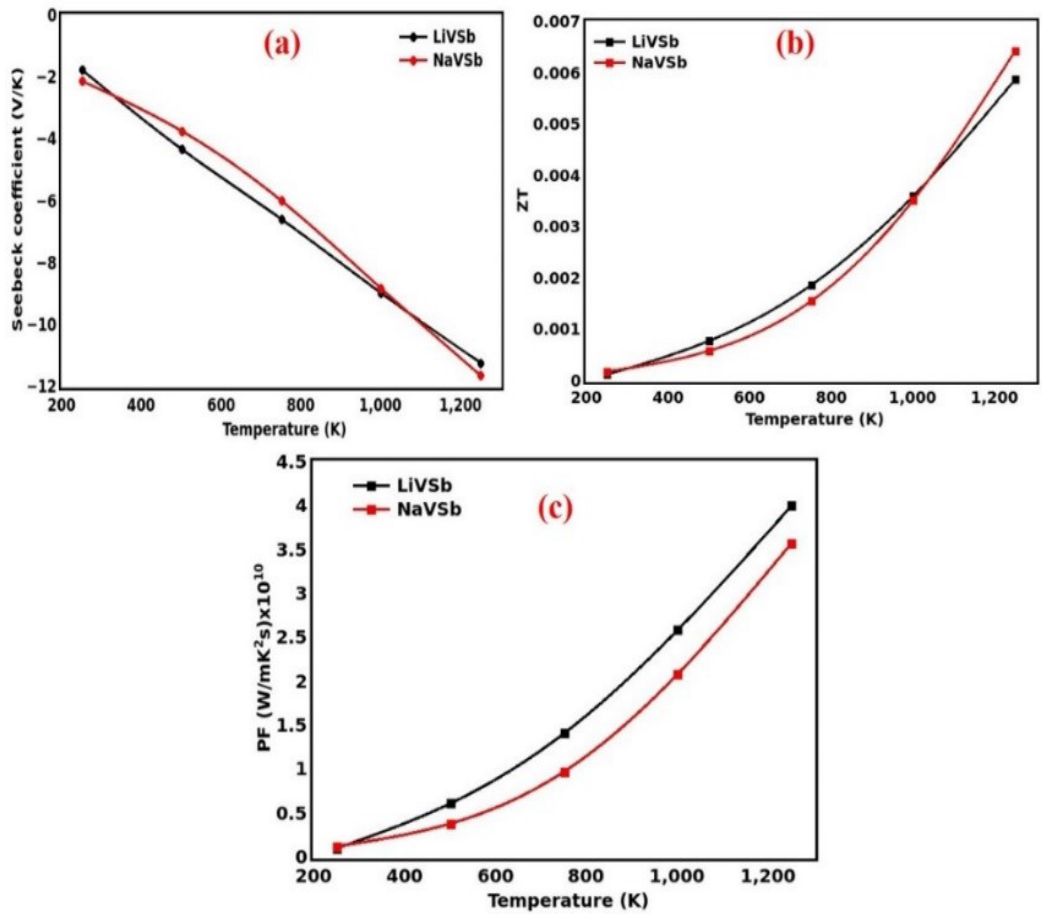

**Fig. 9:** Variation of **(a)** Seebeck coefficient **(b)** figure of merit (zT) and **(c)** power factor (PF) with T for LiVSb and NaVSb alloys

The value of PF and other thermoelectric parameters at different temperature ranges is reported in **Table 5**.

**Table 5:** Calculated value of electrical conductivity (σ/τ in $10^{20}$ Ω ms$^{-1}$), electronic thermal conductivity (σ/τ in $10^{15}$ W/mKs) at 300K and 600K temperature; Seeback coefficient (S in V/K), power factor (PF in $10^{10}$ W/mK$^2$s) and figure of merit (zT) at 300K and 1200 K for LiVSb and NaVSb alloys

| Alloy | σ/τ | | κ/τ | | S | | PF | | zT | |
|---|---|---|---|---|---|---|---|---|---|---|
| | 300K | 600K | 300K | 600K | 300K | 1200K | 300K | 1200K | 300K | 1200K |
| LiVSb | 3.34 | 3.15 | 2.05 | 8.53 | -2.51 | -11.25 | 0.22 | 4.15 | 0.0001 | 0.0060 |
| NaVSb | 2.75 | 2.55 | 1.85 | 6.85 | -1.85 | -11.75 | 0.25 | 3.55 | 0.0001 | 0.0065 |

**Fig. 10 (a)** presents the electrical conductivity as a function of temperature. It is observed that the value of electrical conductivity decreases with increasing temperature. This is due to the reason that with rise in temperature the atomic vibrations rise and leads to large mean free path and more resistance and restrict more electrons to flow consequently there is reduction in the electrical conductivity.

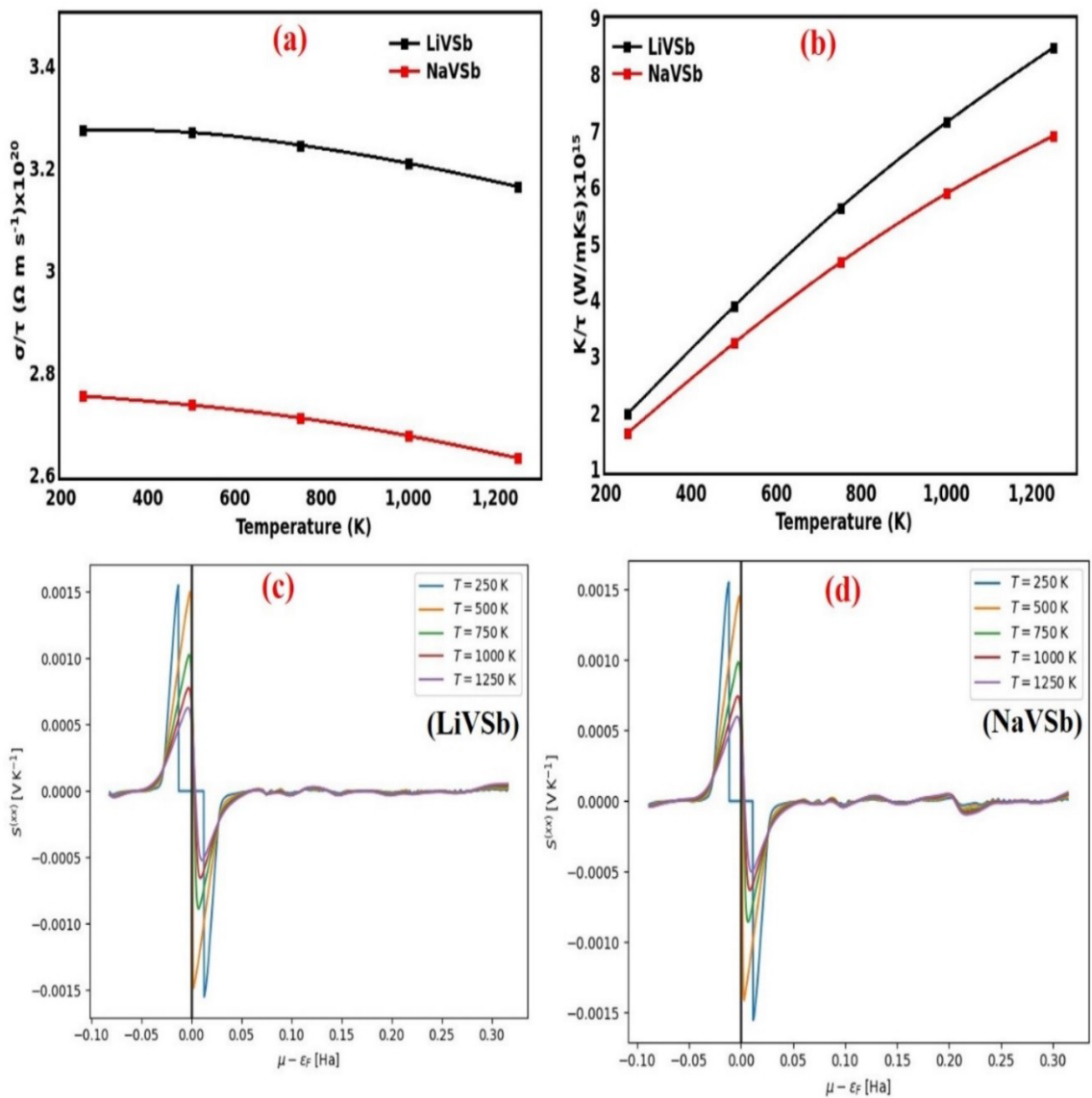

**Fig. 10:** Variation of **(a)** electrical conductivity **(b)** thermal conductivity with T, **(c)** & **(d)** Seebeck coefficient with chemical potential for LiVSb and NaVSb alloys

In **Fig. 10 (b)** we observed a different trend in electronic thermal conductivity with rise in temperature for LiVSb and NaVSb. The calculated values of electrical and electronic thermal conductivity at 300K and at maximum temperature are listed in **Table 5**. In schematic **Fig. 10 (c)** and **(d)** the Seebeck coefficient (S) as a function of chemical potential ) is explored in the range of -0.10eV to 0.30eV to determine the thermoelectric stability. The S is calculated at temperature values; 250K, 500K, 750K, 1000K and 1200K respectively. The maximum and minimum value of S is ±0.0015 for both positive and for negative chemical potential in the locale of µ=0. The observed results adhere to prove LiVSb and NaVSb as ideal thermoelectric materials with robust stability.

### F. Lattice dynamic study

To predict the dynamical behaviour of the crystal lattice phonon band structure are essential to study. With the help of schematic **Fig. 11 (a)** and **(b)** the phonon dispersion curves for LiVSb and NaVSb alloys can be excellently understood. With the reference to the material's crystal structure there are three atoms present in their unit cell which ultimately leads to nine vibrational modes. Among these vibrational modes three are linearly dispersed modes at the Γ point which results from the in-phase atomic displacements of atoms at their basis and called as acoustic modes. The lasting six modes, which exhibit almost flat dispersion near the gamma (Γ) point, arise from the out-of-phase atomic translations and are known as optical modes. In the low-frequency acoustic modes, two transverse modes exhibit degeneracy near the gamma (Γ) point owing to crystal symmetry. Similarly, the crystal symmetry is also evident in the degeneracy of two groups of optical modes near the Γ point. We accurately predict the thermodynamic stability of solids by closely inspecting their phonon band structure. The optical modes, characterized by low dispersion and group velocity, contribute minimally to lattice thermal conductivity however, the major contributors to thermal current are from the acoustic phonons which show high dispersion. Certain materials are prominent for various industrial and commercial applications, like the thermal insulators and thermoelectrics they require very low lattice thermal conductivity for optimal performance.

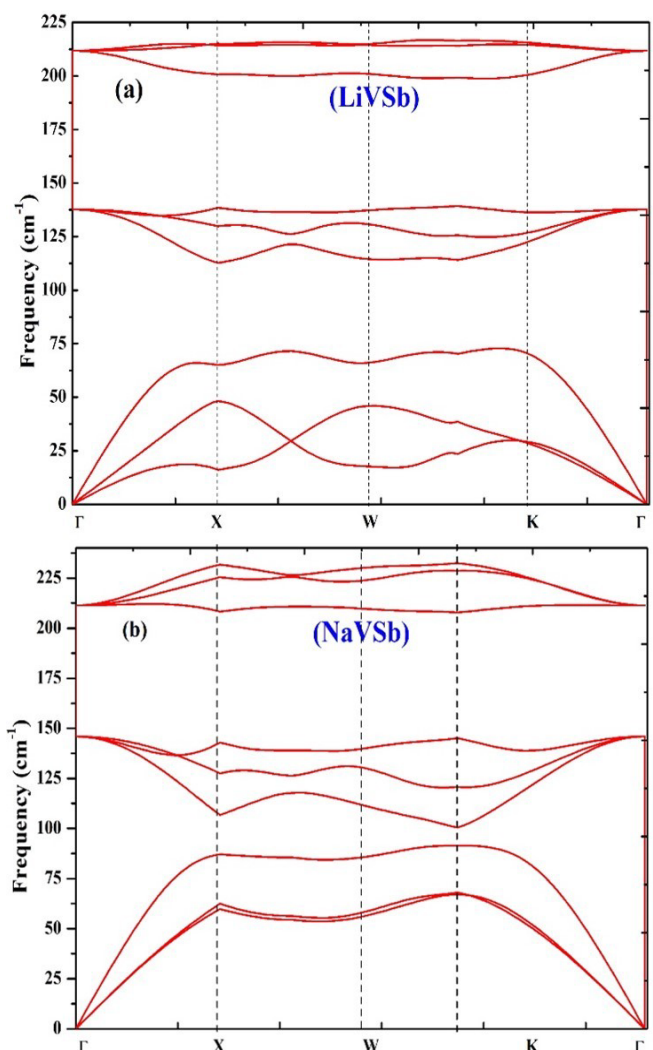

**Fig. 11:** Phonon dynamic stability plots for (a) LiVSb and (b) NaVSb alloys

## IV. Conclusion

In the work reported we have compiled a comprehensive study on LiVSb and NaVSb semi-Heusler alloys, describing their fundamental properties. The DFT calculations approved that these crystalline structures are extremely stable in phase-α and in spin polarised configuration. The electronic band structures profiles reveal the spin-polarized behaviour, signifying the half-metallic property. The band structure divulges that the materials are half-metallic with wide indirect energy gap of 1.40 eV and 1.47eV in spin down channel for both LiVSb and NaVSb alloys respectively. The computed integral values of the magnetic moment are found excellently consistent with Slater Pauling rule. The elasto-mechanical properties, including Pugh's ratio (B/G) divulges the brittle nature of both alloys. Thermodynamic properties determine their stability across a wide range of temperatures at ambient conditions. We evaluated the thermoelectric response, which advocate that the materials are most suitable for converting waste heat energy into useful electrical energy. Additionally, the lattice dynamic investigations confirm the dynamic stability and highly favour the possibilities of their synthesis in future's devices based on renewable energy and solid-state electronics.